\documentclass[%
 reprint,
 amsmath,amssymb,
 aps,nofootinbib
]{revtex4-2}

\usepackage{graphicx}
\usepackage{dcolumn}
\usepackage{bm}
\usepackage{mathtools}
\usepackage{physics}
\usepackage{amsmath}

\newcommand{\nK}{\mbox{ nK }}

\newcommand{\Rb}{\ensuremath{^{87}}Rb}

\newcommand{\eq}[2]{\begin{equation}#1\label{#2}\end{equation}}
\newcommand{\eqal}[2]{\begin{align}#1\label{#2}\end{align}}
\newcommand{\pard}[2]{\frac{\partial #1}{\partial #2}}

\newcommand{\ie}{\textit{i.e. }}

\begin{document}

\preprint{APS/123-QED}

% \title{Comment on Phys.Rev.Lett. 119, 163201 (2017)}
\title{On the absence of the electrostriction force in dilute clouds of cold atoms}

%\author{TBD}

 \author{Arnaud Courvoisier}
 \author{Nir Davidson}
\affiliation{Department of Physics of Complex Systems, Weizmann Institute of Science, Rehovot 761001, Israel}

\date{\today}

\begin{abstract}
The momentum of light in a medium and the mechanisms of momentum transfer between light and dielectrics have long been the topic of controversies and confusion \cite{milonni2010,ll,brevik1978,brevik2018}. We discuss here the problem of momentum transfers that follow the refraction of light by dilute, inhomogeneous ensembles of ultra-cold atoms.
We show experimentally and theoretically that the refraction of light rays by a dilute gas does not entail momentum transfers to first order in the light-atom coupling coefficient, in contradiction with the work reported in Matzliah et al. Phys. Rev. Lett. 119, 189902 (2017).
\end{abstract}

\maketitle
\def\thefootnote{$\dagger$}\footnotetext{Corresponding author : nir.davidson@weizmann.ac.il}
\section{\label{sec:introduction}Introduction}
The fields of optomechanics and collective light scattering by ultra-cold atomic ensembles are active and prolific, as they have seen numerous recent experimental and theoretical developments, on effects such as superradiance \cite{goban2015,araujo2016,roof2016,dimitrova2017}, subradiance \cite{guerin2016} and cooperative shifts \cite{meir2014,ralf2010,peyrot2018}. Most of the recent experimental efforts have been focused on the understanding of the collective response of an ensemble to a probe beam \cite{guerin2017,maiwoger2022}, in the regime of dense atomic samples, in which dipole-dipole interactions are predominant. The work presented here, however, focuses on mechanisms of momentum transfers between light and matter in a regime where ensembles are inhomogeneous and where dipole-dipole interactions are negligible.

We report on new developments concerning electrostriction, a force reported by \textit{Matzliah et. al} in \cite{NH} and thought to emerge when an atomic ensemble acts as a lens and refracts an incoming plane wave. An intuitive momentum conservation picture leads to the derivation of a force acting transversely to the light's propagation direction, and depending linearly on the atomic susceptibility. Our  theoretical and experimental results point towards the non-existence of this force in the form that was previously thought-of. 

We first present a theoretical derivation that predicts the absence of a transverse force resulting from the refraction of a far-off-resonance plane-wave by a dilute atomic ensemble. We then describe a series of well-controlled and well-calibrated experiments aimed at proving the absence of such a force.  

\section{\label{sec:theory}Theoretical considerations}

Well-established theories on the force density experienced by dielectrics in the presence of electromagnetic radiation  predict that, in the limit of non-magnetic dielectric materials where pressure gradients are insignificant, the force density averaged over an optical cycle is given by the following formula \cite{ll,milonni2010,brevik1978}:
\eq{
\vb f = -\frac{\epsilon_0}{2}\vb E^2\grad{\epsilon}\ +\frac{\epsilon_0}{2}\pmb\nabla\left(\pard{\epsilon}{n^*}n^*\vb E^2\right),
}{eq:ll}
where $\epsilon_0$ is the vacuum permittivity, $\epsilon$ is the relative permittivity of the material, $n^*$ is the ensemble density and $\vb E$ is the electric field amplitude. The first term in the force density can be associated to momentum conservation arising from the deviation of light rays by the ensemble, while the second term is the gradient of the interaction energy between the dielectric and the field.

For a dilute gas, $\epsilon = 1+n^*\alpha/\epsilon_0$, where $\alpha$ is the single-atom polarizability. The force density can then simply be re-written as
\eqal{\vb f &= -\alpha\left(\vb\nabla n^*\right)\frac{\vb E^2}{2}+\pmb\nabla\left(\alpha n^*\frac{\vb E^2}{2}\right)\\
&= \frac{n^*\alpha}{2}\pmb\nabla\vb E^2.}{}
The above expression accounts solely for the well-known dipole force, therefore, in the absence of an intensity gradient, one should not measure any back action on the cloud when light propagates through it, to first order in the polarizability. The absence of such a force is due to exact  cancellation between the first and second terms of equation (\ref{eq:ll}) and confirmed by theories that treat this problem in a microscopic framework, based on a dipole-dipole interaction approach \cite{bux2010, bienaime2011, zhu2016, Ayllon2019}. 

Equation (1) of \cite{NH} derived the optomechanical strain resulting from back action to refraction of an incoming plane wave by a dilute atomic ensemble by calculating the change of the electromagnetic momentum  in the eikonal approximation. It captured correctly the transverse parts of the first term of equation (\ref{eq:ll}) but omitted its longitudinal component. Indeed, we note that adding $\Delta p = \hbar k\Delta n $, the Minkowsky correction to the electromagnetic momentum in \cite{NH} naturally yields this missing longitudinal term and therefore turns the prediction of \cite{NH} to be isotropic. 
The second term of equation (\ref{eq:ll}) that does not emerge from momentum conservation considerations was completely neglected in \cite{NH}, leading to a first order force in the single atom polarizability even for plane wave illumination where the dipole force is absent.

\section{\label{sec:exp}Experiment}
We now report on experimental results showing the absence of a back-action following the refraction of light by an atomic ensemble, in the dilute and far-off-resonance limits. The presence of such a force would result in a change in the ensemble's momentum distribution after the passing of a plane wave pulse through the cloud. Therefore, our experimental sequence consists in producing a thermal ensemble at a few hundreds of\nK by means of evaporative cooling in an optical dipole trap, releasing the trap, immediately pulsing a plane-wave on the ensemble for $\tau=400\rm\mu s$, and measuring the cloud's momentum distribution via absorption imaging after 20ms of time-of-flight.  

Our thermal \Rb\ clouds contain approximately $5\times10^5$ atoms at a temperature of 400nK, and we measure the trap oscillation frequency at the end of evaporation to be $\omega \simeq 2\pi\times90\rm Hz$ in all directions. Note that the trap oscillation period is much larger than $\tau$ such that we can neglect thermal motion during the plane-wave pulse.

\begin{figure}[h]
    \centering
    \includegraphics[width = \columnwidth]{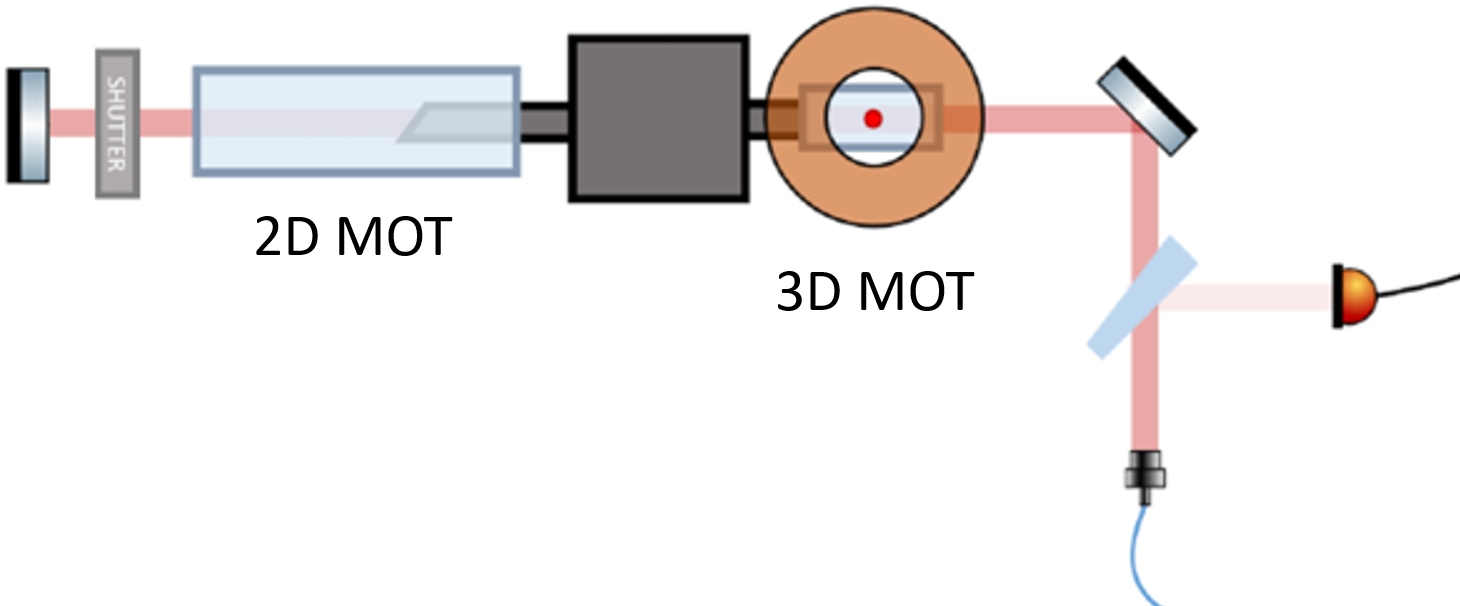}
    \caption{Optical layout used for the electrostriction experiment. The shutter allows us to retro-reflect or not the electrostriction beam onto itself. We monitor the power in the incoming and reflected beams by using a fast-photodiode and a wedged window placed at the fiber's output. The wedged window is used in order to insure that there are no interference fringes on the atoms themselves.}
    \label{fig:es_layout}
\end{figure}

The optical layout used for electrostriction measurements is presented in figure \ref{fig:es_layout}. A beam of waist 1.05mm originates from Thorlabs' F220APC-780 fiber collimator and we check its intensity profile right before the cell using a beam profiler. The beam diameter has been chosen such that the intensity profile impinging on the atoms is quasi-uniform, therefore suppressing the traditional dipole force by more than three orders of magnitude compared to an hypothetical electrostriction force. After entering the vacuum chamber, the beam passes through the ensemble and then propagates along the long axis of the vacuum chamber, from the experiment cell to the 2D MOT cell. This is to ensure that reflections of the beam at glass interfaces will not reach the atoms and cause parasitic dipole forces. The power in the electrostriction beam ranges from 0 to 300mW, and we can continuously vary its detuning with respect to the $F=1$ to $F'=2$ transition of the $D_2$ line of \Rb, in the $-100$GHz to $+100$GHz range.
We added the option of retro-reflecting the beam onto itself in order to create a 1D optical lattice potential. This allowed us to perform precise on-atom light intensity measurements by diffracting the ensemble on the lattice and measuring the oscillations of Kapitza-Dirac orders \cite{Gadway2009}. One can derive a relation between the light intensity, its detuning and the oscillation frequency by numerically solving Mathieu's equation  for a 1D optical lattice \cite{Dalibard2013}. By measuring this oscillation frequency and measuring the laser detuning  with a wave-meter, we directly measure the maximum on-atom intensity to be $4.5\times10^4\rm W.m^{-2}$.
\begin{figure}[h]
    \centering
    \includegraphics[width = \columnwidth]{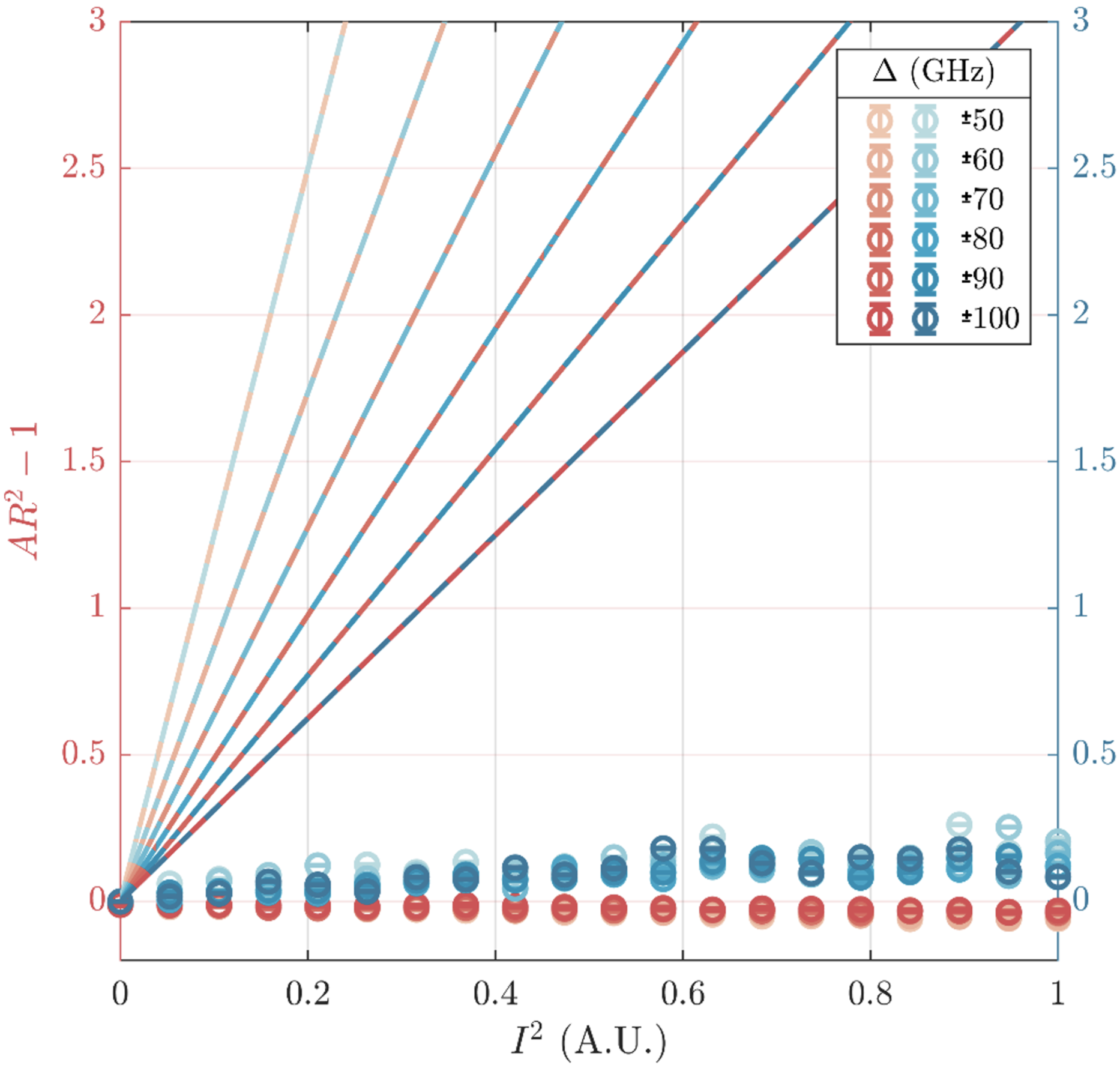}
    \caption{Comparison between our experiment (circular markers) and the theory of \cite{NH} (dashed lines) for detunings between $\pm50GHz$ and $\pm100GHz$, showing the absence of electrostriction as described in \cite{NH}. The aspect ratio, $AR$, was measured after 20ms of time-of-flight. Each data point was obtained by averaging 30 experimental runs and the data was collected in a random order. Error-bars correspond to the standard deviation of the data-spread. The maximum intensity of the incoming beam was measured to be $4.5\times10^4\rm W.m^{-2}$ by retro-reflecting it and measuring the oscillations of Kapitza-Dirac orders \cite{Gadway2009}.}
    %(bottom) Semi-log plot of the dependence of $AR^2-1$ on the detuning, at maximal laser intensity. This shows that the deviation from zero that we observe scales as $1/\Delta$, resulting from a weak parasitic dipole force.}
    \label{fig:noes}
\end{figure}
\vspace{-15pt}
\subsection{Absence of a \textit{transverse} electrostriction force}
In \cite{NH}, \textit{Matzliah et. al} reported on a transverse electrostriction force with respect to the incoming light. They measured variations of the ensemble's aspect ratio after time-of-flight, $AR$, as a function of various experimental parameters such as the laser's detuning and intensity. Note that in the case of a transverse electrostriction force, $AR$ is such that \cite{NH} 
\eq{
AR^2-1=\left(\frac{\sigma_p^{es}}{\sigma_p^{th}}\right)^2,
}{}
where $\sigma_p^{es}$ is the momentum distribution width of the electrostriction impulse, and $\sigma_p^{th}$ is the initial thermal momentum width. Therefore, in the case of a transverse electrostriction force, measuring the cloud's aspect ratio gives a direct measurement of the force. Reproducing the experiment from \cite{NH} in a different experimental setup and in a well-controlled setting, we measured no effect from the electrostriction pulse on the value of $AR^2-1$, in agreement with equation (\ref{eq:ll}) taken in the appropriate limits. In figure \ref{fig:noes}, we show measurements of $AR^2-1$ as a function of the electrostriction beam power and its detuning, $\Delta$, between $\pm50\rm GHz$ and $\pm100\rm GHz$ with respect to the $F=1$ to $F'=2$ transition of the $D_2$ line of \Rb. At detunings of $\pm50\rm GHz$ and at maximal intensity, the theory prediction from \cite{NH} is larger than our measurement by more than a thousand standard deviations. 
%In addition to these new experimental results, we have evidence that the non-unity aspect ratios measured in \cite{NH} were the result of a residual dipole force due to imperfections on the beam used.

When the incoming beam was illuminated through a side window of our 3D MOT cell we measured significant deviations of the aspect ratios from unity, similar to the measurements of \cite{NH}. By carefully monitoring the spatial profile of the beam, we identified fringes across the beam due to weak undesired reflections from the cell walls that are unavoidable in the case of side illumination. We thus asses that the aspect ratios measured in \cite {NH} resulted from a residual dipole force due to similar imperfections on the beam used.     

\subsection{Absence of an \textit{isotropic} electrostriction force}
In addition to the absence of a transverse force, we show experimentally that, in the case of plane wave illumination and in the far-off-resonance limit, the two terms in equation (\ref{eq:ll}) compensate one-another. To that end, we systematically measured the width of the ensemble's momentum distribution with and without pulsing the electrostriction beam. Initially, our thermal ensembles exhibit Gaussian position and momentum distributions with 
\eq{
\sigma_r^{th} = \sqrt{\frac{k_BT}{m\omega^2}}\text{\ and \ }\sigma_p^{th} = \sqrt{k_BTm}.
}{}
We define $\beta$ as the ratio between the Gaussian width of the cloud after time-of-flight in the pulsed and un-pulsed cases. Upon acting on the ensemble, $\vb f$ changes the momentum distribution of the cloud to
\eq{
\sigma_p^{pulsed}=\sqrt{{\sigma_p^{th}}^2+{\sigma_p^{\vb f}}^2}.
}{}
Similarly to the derivation shown in \cite{NH}, after a time-of-flight long with respect to $2\pi/\omega$, the ratio $\beta$ is such that 
\eq{
\beta^2-1=\left(\frac{\sigma_p^{\vb f}}{\sigma_p^{th}}\right)^2.
}{}
\begin{figure}[h]
    \centering
    \includegraphics[width = \columnwidth]{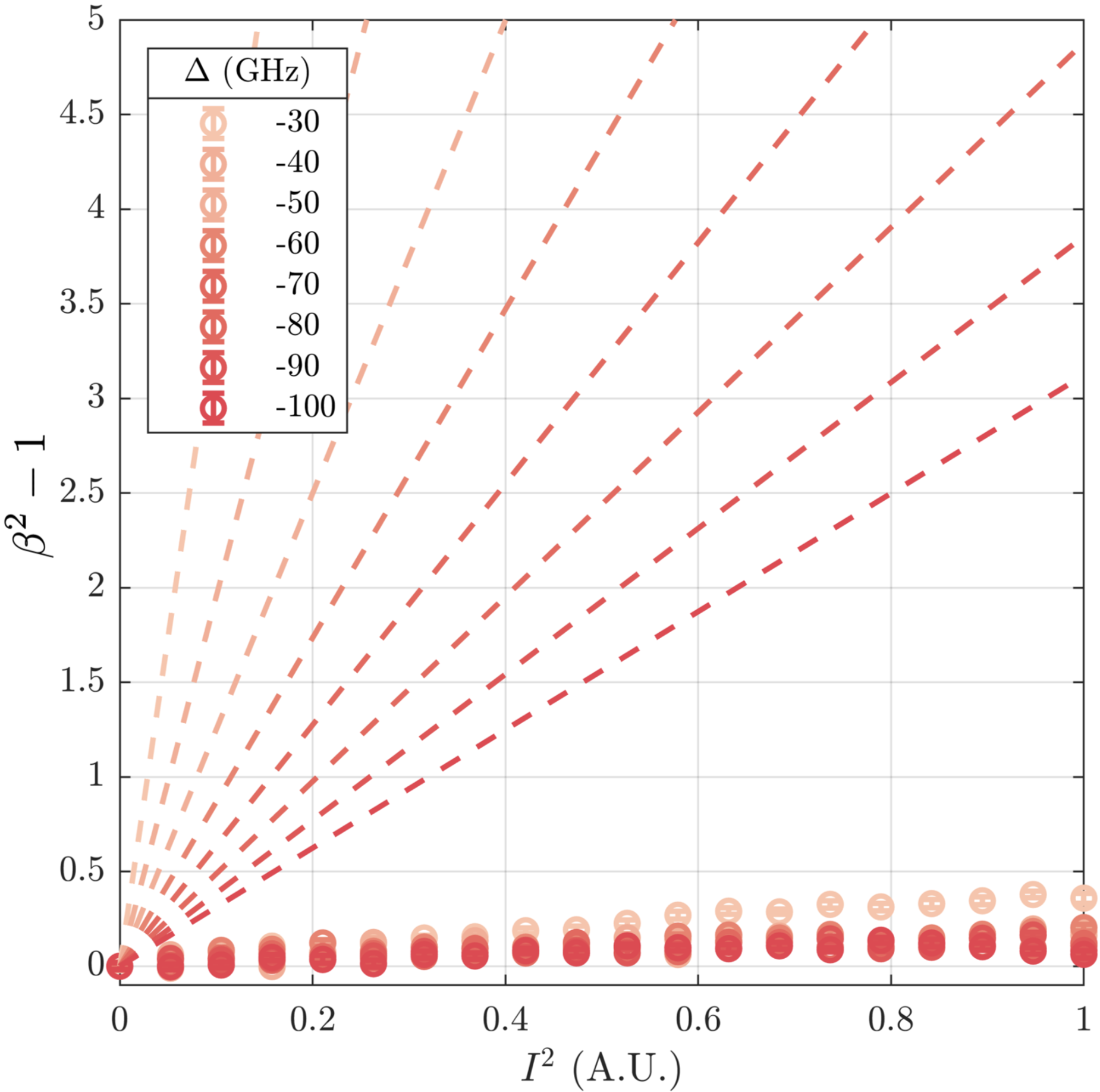}
    \caption{Comparison between our experiment (circular markers) and an incomplete theory where only the first term of equation (\ref{eq:ll}) would be present (dashed lines), for detunings between $-30GHz$ and $-100GHz$. The $\beta$ parameter was measured after 20ms of time-of-flight. Each data point was obtained by averaging 30 experimental runs and the data was collected in a random order. Error-bars correspond to the standard deviation of the data-spread. The maximum intensity of the incoming beam was measured to be $4.5\times10^4\rm W.m^{-2}$ by retro-reflecting it and measuring the oscillations of Kapitza-Dirac orders \cite{Gadway2009}.}
    %(bottom) Semi-log plot of the dependence of $AR^2-1$ on the detuning, at maximal laser intensity. This shows that the deviation from zero that we observe scales as $1/\Delta$, resulting from a weak parasitic dipole force.}
    \label{fig:noforce}
\end{figure}
Therefore, measuring $\beta^2-1$ yields a direct measurement of the action of $\vb f$ on the ensemble. In figure 3, we show the evolution of $\beta^2-1$ as a function of the electrostriction beam power for various detunings as well as the theoretical expectations in the case where only the first term of equation (\ref{eq:ll}) would be present \ie if we were to follow the intuitive picture of momentum conservation that follows the refraction of light rays by the cloud.

As in the transverse case, at maximal intensity and at a detuning of $-30\rm GHz$, the measured value of $\beta^2-1$ and the one predicted by the incomplete theory differ by more than a thousand standard deviations. We attribute the slight deviation from zero to incoherent scattering by the electrostriction beam. We can conclude on the absence of a mechanical back-action on a dilute atomic ensemble through which a plane wave propagates, to first order in the atomic susceptibility. 
\vspace{10pt}
\section{Conclusion} It is clear from the theoretical picture we presented that one should not expect a mechanical back action to first order in $\alpha$ to the refraction of light by a dilute atomic ensemble. The deflection of light rays imparts a momentum kick to a volume element of the cloud which, alone, would lead to a decrease of the ensemble's density in the case of a red-detuned beam. However, locally, the interaction energy between the medium and the electric field would increase if the density was to get lower. There is therefore a force that prevents this from happening. In the dilute limit \ie when the susceptibility is proportional to the density, these two forces compensate one another exactly. Outside of the dilute regime, the interaction energy and the momentum conservation condition have a different dependence on the atomic density, such that the two forces do not compensate one another, and one can derive forces proportional to $\alpha^2$ \cite{maiwoger2022,Ayllon2019}. 

We have described above our recent experimental efforts where careful beam calibration and characterization reveal the absence of electrostriction as reported in \cite{NH}. We believe that the results previously reported were the result of a parasitic dipole force due to imperfections of the illuminating beam and do not reflect a new physical mechanism. These experimental results are in agreement with well-established theories that predicts exact cancellation between the two terms of equation (\ref{eq:ll}) and hence the absence of an electrostriction force in the dilute regime.

\section*{Acknowledgements}
The authors acknowledge fruitful discussions with Ulf Leonhardt and Ephraim Shahmoon and thank Amruta Gadge for technical assitance.
\bibliography{biblio}

\end{document}